\documentclass{aa}
\usepackage{graphics}
\begin{document}

   \thesaurus{ ?    % A&A Section ?
              (13.18.1;  % Radio continuum: galaxies,
               11.01.2;  % Galaxies: active,
               11.03.2;  % Galaxies: compact 
               11.10.1;  % Galaxies: jets 
               11.14.1;  % Galaxies: nuclei 
               11.17.4 3C~138)} % Quasars: individual: 3C~138

   \title{Superluminal motion in a compact steep spectrum 
          radio source 3C~138}

%   \subtitle{I. Overviewing the $\kappa$-mechanism}

   \author{Zhi-Qiang Shen\inst{1},
           D. R. Jiang\inst{2},
           Seiji Kameno\inst{3},
           \and
           Y. J. Chen\inst{2}
          }

   \offprints{Z.-Q. Shen }

   \institute{Institute of Space and Astronautical Science, Yoshinodai 3-1-1, 
              Sagamihara, Kanagawa 229-8510, Japan\\
              email: zshen@vsop.isas.ac.jp
         \and
             Shanghai Observatory, Shanghai 200030, China\\
              email: djiang, cyj@center.shao.ac.cn
         \and
             National Astronomical Observatory, Osawa 2-21-1, Mitaka, 
             Tokyo 181-8588, Japan\\
             email: kameno@hotaka.mtk.nao.ac.jp
             }

   \date{Received , 2000; accepted , 2000}

 \titlerunning{The center of activity in 3C~138}

 \authorrunning{Z.-Q. Shen et~al.}

  \titlerunning{the center of activity in a CSS radio source 3C~138}

   \maketitle

   \begin{abstract}

   We present the results of 5~GHz VLBI observations 
   of a compact steep spectrum source 3C~138. 
   The data are consistent with
   the western end being the location of the central activity.
   The observed offset between different frequencies in the central region
   of 3C~138 can be accounted for by a frequency dependent shift of 
   the synchrotron self-absorbed core.
   Our new measurements confirm the existence of a superluminal motion,
   but its apparent velocity of 3.3c is three times slower than the reported 
   one. This value is consistent with the absence of parsec-scale counter-jet
   emission in the inner region, but seems still too high to allow 
   the overall counter-jet to be seen in terms of Doppler 
   boosting of an intrinsically identical jet.   
   Either an interaction of
   jet with central dense medium, or an intrinsically 
   asymmetrical jet must be invoked to reconcile the detected
   superluminal speed with the observed large scale asymmetry
   in 3C~138.

   \end{abstract}

%
%________________________________________________________________

\section{Introduction}

   The radio source 3C~138 (=~4C~16.12~=~J0521+1638)
   is identified with a quasar with {\emph m$_v$}=18.84 and 
   {\emph z}=0.759 (Hewitt \& Burbidge \cite{hewitt}). It is a compact,
   powerful quasar with a well-defined turn-over in the spectrum 
   at about 100~MHz and a steep high frequency spectrum of 0.65 
   (S~$\propto$~$\nu^{-\alpha}$), making it a prototype of the compact 
   steep spectrum (CSS) radio sources (Fanti et~al. \cite{fanti90}).
   Its spectrum remains straight up to 22~GHz, supporting that no
   dominant flat spectrum core exists (Kameno et~al. \cite{kameno}).

   The radio structure of 3C~138 consists of a core with a bright jet
   and compact lobe to the east and a fainter, more diffuse lobe to the
   west from the VLA and MERLIN observations (Akujor et~al. \cite{akujor} and 
   references therein; L\"{u}dke et~al. \cite{ludke}). High resolution VLBI
   observations (Geldzahler et al. \cite{geldzahler}; Fanti et al. \cite{fanti89}; 
   Nan et al. \cite{nan}; Dallacasa et~al. \cite{dallacasa}; Cotton et al. 
   \cite{cotton}) reveal several knots within 
   the main jet emission extending about 400~mas in a position angle of 
   65$^\circ$ and a low brightness counter-jet about 250~mas away in the 
   opposite direction. There are two knots dominating the central 10~mas region. 
   Fanti et al. (\cite{fanti89}) claimed on the basis of apparent size that 
   the eastern knot is the core. Cotton et al. (\cite{cotton}) argued that 
   the western one is the location of the
   nuclear core mainly because of its very weak polarization ($<0.4\%$)
   compared to a peak polarized intensity of $3.5\%$ for the eastern
   component at 5.0~GHz. Cotton et~al. (\cite{cotton}) also suggested
   a relative proper motion of 0.27~mas~yr$^{-1}$ between the two knots,
   corresponding to an apparent velocity of 9.7c 
   (assuming H$_0$~=~65~km~s$^{-1}$~Mpc$^{-1}$ and q$_0$~=~0.5, hereafter).
           
   In general, the radio emission from CSS sources is thought to be
   relatively free of beaming effects, consistent with the fact that
   CSS sources are the least variable of compact extragalactic radio 
   sources, and CSS sources are rarely found to be superluminal (O'Dea 
   \cite{odea}). When combined with the observed two-sided jet structure, 
   such a superluminal motion in 3C~138, if confirmed, will strictly 
   confine the jet velocity and its direction to the line of sight of the 
   observer. We report on 5.0~GHz VLBI observations of 3C~138 
   as an attempt to confirm its proper motion and probe its compact 
   radio core area.
   
%__________________________________________________________________

\section{Observations and data reduction}

%                                     Two column figure (place early!)
%______________________________________________ Gamma_1 (lg rho, lg e)
   \begin{figure*}
     \vspace{-1.0cm}
      \resizebox{\hsize}{!}{\includegraphics{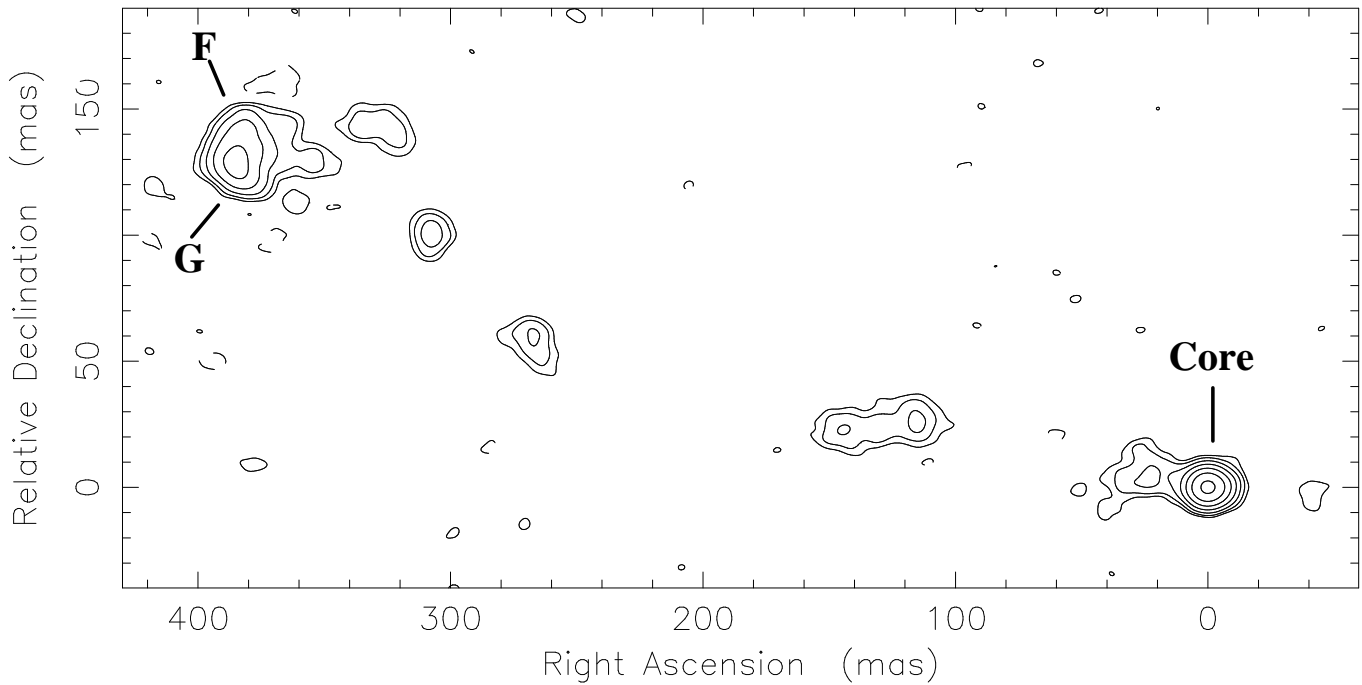}}
%\rule{0.4pt}{4cm}% line thickness, height of picture
\vspace*{-2.5cm}
     \caption{Tapered 5~GHz VLBI image of 3C~138 restored with a 10~mas 
              circular beam. Contour levels are drawn at $-$3.6, 
              3.6$\times$2$^n$~mJy~beam$^{-1}$ (n~=~0, 1, ..., 6). 
              The peak flux density is 274~mJy~beam$^{-1}$. The RMS noise 
              is 1.2~mJy~beam$^{-1}$.
%              }
\label{Fig1}}%
    \end{figure*}

   The VLBI observations were carried out at 5.0~GHz on November 8, 1997
   with the enlarged European VLBI Network composed of Effelsberg (100-m), 
   Hartebeesthoek (26-m), Jodrell MK2 (25-m), Medicina (32-m), Noto (32-m), 
   Onsala (25-m), Shanghai (25-m), Torun (32-m), Urumqi (25-m) and Westerbork 
   (25-m). Sources were observed in a snapshot mode, i.e., 5-12 thirteen-minute
   scans spreading across a wide range in hour angle to obtain a good u-v
   coverage. The left-circular polarized radio signals were recorded in 
   Mark~III mode C with a total bandwidth of 14~MHz (7 IFs and 2~MHz per 
   IF band) at each station. The correlation of the data was performed on
   the MPIfR MK~III Correlator in Bonn, Germany with an output averaging 
   time of 4~s.

   Then, {\sl a priori} visibility amplitude calibrations for each antenna 
   were applied using the antenna gain and the system temperatures measured 
   during the observations. After the data were globally fringe-fitted 
   in AIPS (Schwab \& Cotton \cite{schwab}), the subsequent data inspection, 
   imaging
   and modelfitting were carried out with DIFMAP (Shepherd \cite{shepherd}).
   
   Because of the large separation ($\sim$400~mas) between the eastern compact 
   lobe and central core region in 3C~138, special attention was paid to the 
   smearing effects of time averaging and bandwidth in order to image an 
   undistorted full field. The 2~MHz data in each IF were not averaged 
   through 7 IFs, and the averaging time was only 12~s to minimize smearing 
   effects.
   
   Images were produced with the usual iterations of cleaning and phase 
   self-calibration. Only a constant gain factor for each IF band of each 
   antenna was implemented in the later stage of imaging. These gain corrections
   are consistent with those obtained for other sources (Shen et~al. 
   in preparation). To improve the sensitivity 
   to the low surface brightness regions, the visibility data were naturally 
   weighted and tapered with FWHM~=~30M$\lambda$. 

\section{Results}

\subsection{Overall structure}

   The overall morphology of 3C~138, as shown in 
   Fig.~\ref{Fig1}, resembles past VLBI 
   images. There are two notable emission regions: the central core and the
   eastern lobe, which are separated by 400~mas at a position angle of 
   70$^\circ$. In between them are several discrete emission components.
   
   The total cleaned flux density in the image (Fig.~\ref{Fig1}) is $\sim$0.85~Jy, 
   which is the same as the correlated flux density on the shortest baseline, 
   Westerbork-Jodrell MK2, but far less than the values of 3.24~Jy and 3.58~Jy 
   by Dallacasa et al. (\cite{dallacasa}) and Cotton et al. (\cite{cotton}), 
   respectively. The single-dish measurements from the University of Michigan 
   Radio Astronomy Observatory (UMRAO) show a very stable total flux density
   of 3.80~Jy at 4.8~GHz with a variation less than 6\% over past 10 years.
   The source is heavily resolved on baselines longer 
   than 3~M$\lambda$ (see Fig.~2c in Fanti et al. \cite{fanti89}).
   So, the main reason for the shortage of cleaned flux density in our image 
   is the lack of shorter baselines like MERLIN or VLA-Pie Town (cf. Dallacasa 
   et al. \cite{dallacasa}). As a result, the more extended, lower surface 
   brightness regions seen in previous images (e.g. Fanti et~al. \cite{fanti89}) 
   are not well 
   recovered in our image, which causes discontinuity in the main jet emission 
   towards the eastern lobe. The connection of separated emissions (including 
   the core region), however, exhibits a sign of oscillation seen from other 
   observations. No counter-jet was significantly detected in our image though 
   there are some very weak diffuse emissions at about 3 times noise 
   level around the counter-jet region of the 1.7~GHz image by Cotton et~al. 
   (\cite{cotton}). The RMS noise in the image is 1.2~mJy/beam, about twice
   the expected thermal noise for the total on-source time of 1.3~hr. 

   Two compact components (labeled F and G after Fanti et~al. \cite{fanti89}) 
   in the eastern lobe can be clearly seen. They are aligned nearly orthogonal 
   to the jet axis, which may be interpreted as a working surface  of the jet
   due to its interaction with the surrounding medium.
  
\subsection{Central core region}    

%
%                                                One column figure
%----------------------------------------------------------- S_vib
   \begin{figure}
%      \vspace{5cm}
 \resizebox{\hsize}{!}{\includegraphics{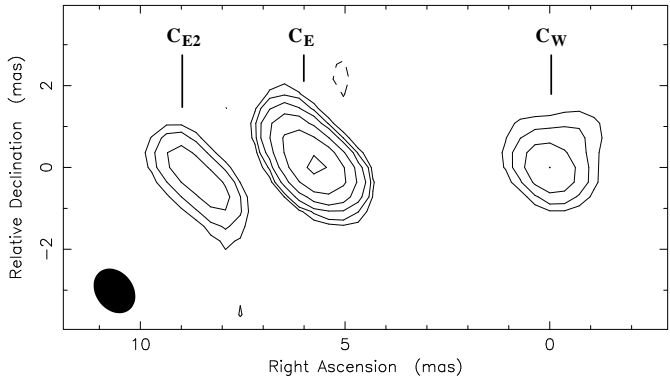}}
\vspace*{-1.5cm}
      \caption[]{Full resolution 5~GHz VLBI image of the central core region 
                 of 3C~138. The resolution is 1.2~mas$\times$~0.92~mas at a 
                 position angle of 37$^\circ$ (indicated at the lower left 
                 corner). Contour levels are drawn at $-$3.6, 
                 3.6$\times$2$^n$~mJy~beam$^{-1}$ (n~=~0, 1, ..., 5).
                 The peak flux density is 134~mJy~beam$^{-1}$. The RMS noise 
                 is 1.2~mJy~beam$^{-1}$.
%              }
              \label{Fig2}}
   \end{figure}

   For the purpose of identifying the center of activity, a full resolution 
   image is displayed in Fig.~\ref{Fig2}. This image was made from the same calibrated
   data as Fig.~\ref{Fig1}, except that the baselines shorter than 30~M$\lambda$ were 
   removed and the uniform weighting was adopted to ensure a high resolution. 
   To yield a quantitative description, modelfitting to both amplitudes and 
   phases in the calibrated visibility data was applied. Because of the
   complicated overall structure and the limited observational data, we fit
   a model to core region only while keeping the cleaned components for the 
   remaining structure (cf. Piner et~al. \cite{piner}). In this way, we can
   minimize the effect mainly from the components F and G. We also tried a
   modelfitting to the longer baselines data corresponding to Fig.~\ref{Fig2}. We even
   made an image of the central region only with a 1-minute averaging time 
   and then fit the visibility data. All these results are quite consistent.
   In Table~1 is presented the best fitting model consisting of three circular 
   Gaussian components. 
   
   The sizes of components C$_E$ and C$_W$ are about the same, but C$_E$ 
   is much brighter. The flux density of C$_E$ is stable over past 12 years
   (cf. Akujor et~al. \cite{akujor}; Dallacasa et~al. \cite{dallacasa}).
   The two point spectral indices ($\mathrm{\alpha_{5.0GHz}^{8.6GHz}}$) 
   obtained by combining the high resolution VLBA results at 8.6~GHz (Fey \& 
   Charlot \cite{fey}) are almost the same for both C$_E$ and C$_W$ (see 
   Table~1).  Component C$_{E2}$ is less well constrained because it is much 
   too resolved. Actually, C$_{E2}$ emission has appeared in several previous 
   5~GHz images (e.g. Fig.~13 in Cotton et~al. (\cite{cotton})), but was first 
   fitted at 8.6~GHz by Fey \& Charlot (\cite{fey}).
   
%   These properties
%   make the identification of the real VLBI core uncertain, which will be 
%   discussed in Sect.~4. 
 
%__________________________________________________ One column table
   \begin{table}
      \caption[]{Gaussian model for the central region of 3C~138}
      \[
         \begin{array}{p{0.20\linewidth}rrrcc}
            \hline
            \noalign{\smallskip}
 Component & \mathrm{S}/\mathrm{[Jy]} & \mathrm{r}/\mathrm{[mas]} 
           & \mathrm{\theta}/\mathrm{[^\circ]} & \mathrm{a}/\mathrm{[mas]} 
	 & \mathrm{\alpha_{5.0GHz}^{8.6GHz} (S\propto\nu^{-\alpha})} \\
%       & \mathrm{(Jy)} & \mathrm{(mas)}  & \mathrm{(^\circ)} &  \mathrm{(mas)} 
%         &  &  \mathrm{(^\circ)} \\
                 \noalign{\smallskip}
            \hline
            \noalign{\smallskip}
C$_E$ \dotfill & 0.218 &      -5.76        &  90.0        & 0.768 & 0.46  \\
C$_{E2}$ \dotfill & 0.051 & -8.75 & 91.7 & 1.320 & 0.45 \\
C$_W$ \dotfill & 0.051 & 0 & 0 & 0.753 & 0.45 \\ 
            \noalign{\smallskip}
            \hline
         \end{array}
      \]
S: the flux density in Jy; (r, $\theta$): the distance and position angle 
of each component with respect to component C$_W$ 
in mas and degrees, respectively; a: the diameter (FWHM) of circular Gaussian 
component in mas; $\mathrm{\alpha_{5.0GHz}^{8.6GHz}}$: two point spectral index.
%$\mathrm{\mu}$: proper motion relative to C$_W$ in $\mathrm{(mas~yr^{-1})}$.
   \end{table}
\section{Discussion}

\subsection{Proper motion}    

%
%                                                One column figure
%----------------------------------------------------------- S_vib
   \begin{figure}
%      \vspace{-5cm}
 \resizebox{\hsize}{!}{\includegraphics{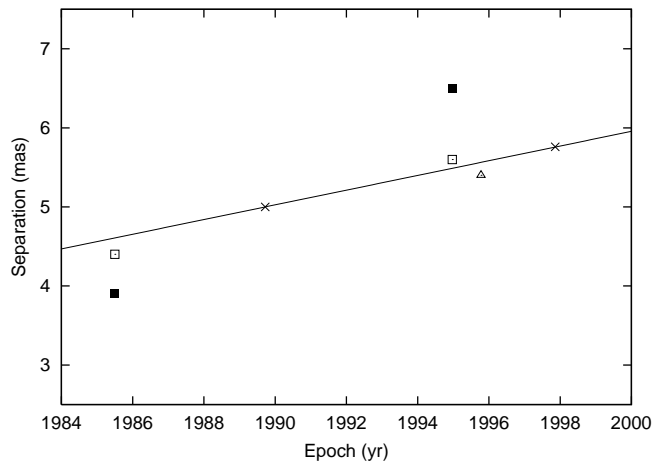}}
%\vspace*{-1.5cm}
      \caption[]{The apparent separation of two central components
                 C$_W$ and C$_E$ in 3C~138 as a function of the
                 observing epoch. These measurements are made at
                 5.0~GHz except one (denoted by a open triangle)
                 at 8.6~GHz. Symbols are explained in the text.
                 Also shown is a fitting line (see text).
%              }
              \label{Fig3}}
   \end{figure}

   There are three published 5~GHz VLBI measurements made at epochs 1985.5 
   (Fanti et~al. \cite{fanti89}), 1989.72 (Dallacasa et~al. \cite{dallacasa})
   and 1994.97 (Cotton et~al. \cite{cotton}). Fanti et~al. (\cite{fanti89})
   resolved the central region with two compact components, but did not
   present the separation at the time. Dallacasa et~al. (\cite{dallacasa}) 
   made the first measurement of the apparent separation of 5.0~mas. 
   Cotton et~al. (\cite{cotton}) estimated the separation of 3.9 and 6.5~mas 
   for epochs 
   1985.5 and 1994.97 respectively, and obtained an averaged rate of 
   separation of 0.27~mas~yr$^{-1}$ from all three-epoch observations 
   (see Table~1 and Fig.~14 of Cotton et~al. (\cite{cotton})). A separation 
   of 6.5~mas at epoch 1994.97 is larger than that of 5.76~mas at a later epoch
   1997.86 in Table~1. This would be indicative of a back flow in the jet 
   motion, which is very unlikely if not impossible. The 8.6~GHz VLBA 
   observations (Fey \& Charlot \cite{fey}) measured a separation of only 
   5.4~mas at epoch 1995.78, which is also
   much less than that at epoch 1994.97. To further investigate
   this, we measured the peak-to-peak separation from images directly,
   and then compared them with those modelfitting results. As a result,
   we got an excellent agreement between the modelfitted separation and 
   the peak-to-peak separation at epochs 1989.72 and 1997.86 at 5.0~GHz
   and at epoch 1995.78 at 8.6~GHz, but the discrepancy at epochs 1985.5 
   and 1994.97 was as large as 0.5 and 0.9~mas, respectively. This difference
   is about one-half and one-quarter of the corresponding beam size, and
   quite difficult to understand for us at present.  
   
   Therefore, we decided to estimate the proper motion by using two epoch
   (1989.72 and 1997.86) 5~GHz data (denoted by crosses in
   Fig.~\ref{Fig3}) only. As shown in Fig.~\ref{Fig3}, this produces a 
   relative proper motion of 0.093~mas~yr$^{-1}$ between C$_E$ and C$_W$,  
   which corresponds to a superluminal speed of 3.3c. We used other two
   5.0~GHz observations to evaluate its uncertainty. The peak-to-peak 
   separation (denoted by open squares in Fig.~\ref{Fig3}) of 4.4 and 5.6~mas 
   at the corresponding epochs 1985.5 and 1994.97 gave an uncertainty
   of 5\%, while the published points (denoted by filled squares in 
   Fig.~\ref{Fig3}) would introduce an uncertainty of 16\%. This  
   proper motion is much slower than the reported 0.27~mas~yr$^{-1}$ 
   (Cotton et~al. \cite{cotton}). The ongoing monitoring observations 
   confirm that the knot is definitely moving much slower than suggested
   by the earlier data (Bill Cotton, private communication 2000). 

\subsection{Synchrotron self-absorption} 

   There is a relative position offset of 0.17~mas at epoch 1995.78
   for C$_E$ between the position interpolated from 5~GHz data (fitting
   line in Fig.~\ref{Fig3}) and the measurement made at 8.6~GHz (Fey \& 
   Charlot \cite{fey}) denoted by a open triangle in Fig.~\ref{Fig3}. 
   Such a frequency dependence of the observed position can be 
   explained by synchrotron self-absorption (Lobanov \cite{lobanov}).
   To estimate this position offset using eq. (11) of Lobanov 
   (\cite{lobanov}), we have searched the literature for its radio
   spectral energy distribution. As a result, the total synchrotron 
   luminosity of 1.5$\times$10$^{45}$~erg~s$^{-1}$ is estimated. A 
   superluminal speed of 3.3c implies a maximum angle $\theta$ 
   to the line of sight, a minimum jet speed $v~(= \beta $c) and a minimum 
   Lorentz factor $\gamma$ of 34$^\circ$, 0.96c and 3.5, respectively.
   The projected opening angle $\phi_\mathrm{o}$ of 5.8$^\circ$ is measured from 
   C$_E$ component size over its distance to C$_W$ at 8.6~GHz. These values
   predict a maximum position offset of 0.22~mas between 5.0 and 8.6~GHz 
   under the assumption of energy equipartition between radiating particles 
   and magnetic field. The upper limit is mainly due to the uncertainty in 
   the determination of 
   the viewing angle $\theta$ and the resultant lower limits to both 
   $\beta$ and $\gamma$. Thus, such an observed offset in the central region
   of 3C~138 can be well explained as a frequency dependent shift of the
   synchrotron self-absorbed core. Assuming a conical jet geometry in
   which the magnetic field decreases linearly with its distance to the core, 
   we can further derive the magnetic field in the 5~GHz VLBI core region of 
   0.12, 0.02 and 0.01~G for components C$_W$, C$_E$ and C$_{E2}$, 
   respectively (cf. Lobanov \cite{lobanov}).
   
   Recently, a significant position offset has been reported for another 
   superluminal CSS source 3C~216 (see Fig.~3 of Paragi et~al. 
   \cite{paragi00}). These may suggest that the core region in CSS sources 
   could also be described by ultracompact jet models, which have been
   successful in interpreting the central region of radio-loud active 
   galactic nuclei (cf. Lobanov \cite{lobanov} and references therein).
      
   We are aware, however, that this offset may be due to the positional
   uncertainties of measurements and limited resolution and blending at
   lower frequency. Future high resolution, multi-frequency VLBI 
   observations will be critical to clarify this.     

\subsection{Core identification}
   
   It is hard to determine the real core in 3C~138 based on the apparent 
   source size and spectral index available now. From their high resolution 
   polarization sensitive imaging, Cotton et~al. (\cite{cotton}) have ruled 
   out the possibility that C$_E$ was the core and further suggested that 
   C$_W$ harbors the central energy source. The non variation in the flux 
   density of C$_E$, discussed in Sect.~3.2, can also be viewed as an 
   indication that C$_E$ probably is not related to the center of activity.
   Future observations with better resolution at lower frequencies (such as
   the 1.6~GHz space VLBI (VSOP) imaging observations (cf. Hirabayashi et~al.
   \cite{hirabayashi98})) and better sensitivity
   at higher frequencies should be able to clearly separate and detect the
   two compact components C$_W$ and C$_E$ and, thus enable us to 
   undoubtedly determine if the spectral index, as suggested by Cotton et~al.
   (\cite{cotton}), is flatter in component C$_W$.
   
   Our new measurements confirm the systematic increasing in the knot 
   separation with time. But the rate is three times slower than
   the previous one. The detected superluminal speed, when combined with
   the observed two-sided jet structure in 3C~138, poses very strong
   constraints on the jet kinematics. 
   
   The expected ratio of the flux density between jet and counter-jet 
   is given by  
   $\mathrm{R} = (\frac{1 + \beta \mathrm{cos}\theta}
       {1 - \beta \mathrm{cos}\theta}$)$^{2+\alpha}$.
   Here, intrinsically symmetrical two-sided jets with the
   same speed in two opposite directions are assumed.

   The ratio of the flux density of the eastern to western lobe seen in
   the 1.7~GHz image (Cotton et~al. \cite{cotton}) is about 16. Using the 
   spectral index between 1.7 and 5.0~GHz of 0.75 (Fanti et~al. 
   \cite{fanti89}), we obtained the product $\beta \mathrm{cos}\theta = 0.46$, 
   in good agreement with the value
   estimated at 5~GHz 
% flux density ratio using the MERLIN+EVN data
   (Fanti et~al. \cite{fanti89}). But this restricts to a maximum apparent
   speed of 1.6c, which is significantly smaller than 3.3c. 
   Therefore, the superluminal motion of 3.3c in the central core region,
   though much less than earlier reported one, is still too high to
   allow the overall counter-jet structure to
   be seen in terms of the projection effect and Doppler boosting of
   an intrinsically identical jet. 
   
   One solution to circumvent this is to assume a non-uniform distribution of 
   R for the whole jet since usually lobes are observed to move at a much 
   lower speed. This can naturally explain the absence of parsec-scale 
   counter-jet in the central region. Cotton et~al. (\cite{cotton}) 
   obtained a lower limit to the ratio R of 20 within the central area. 
   Our new observations slightly raised this lower limit to 30 at 5 times 
   RMS noise level. These lower limits are consistent with the detected 
   superluminal motion in the inner region being due to the relativistic 
   boosting in the jet.
   
   Such a change in R could be caused by the strong interaction of the jet 
   with the surrounding medium in the immediate region of the core.
   This is supported by the existence of a 
   large amount of ionized gas in the central region (Cotton et~al. 
   \cite{cotton}). The bending in the jet or a helical jet model is a
   possible explanation, but unlikely to affect R significantly in case
   of 3C~138 (cf. Fanti et al. \cite{fanti89}). A superluminal speed of 
   3.3c implies a ratio R greater than 360 (using the spectral index of 
   component C$_E$ in Table~1), 12 times the lower limit estimated above.
   On the other hand, the possibility that 
   the observed overall asymmetry is intrinsic is still there, but difficult
   to test because of the limited data gathered. The nature of the core region 
   is worthy of a more detailed study.

\begin{acknowledgements}
We wish to thank Carla Fanti, the referee, for a careful reading of
the manuscript and very useful comments, and Bill Cotton for providing 
us with the 1.7~GHz VLBI 
image and for communication of information prior to publication.
This research has made use of the NASA/IPAC Extragalactic Database (NED)
which is operated by the Jet Propulsion Laboratory, Caltech, under
contract with the National Aeronautics and Space Administration.
This research has made use of data from the University of Michigan
Radio Astronomy Observatory which is supported by the National 
Science Foundation and by funds from the University of Michigan.

\end{acknowledgements}

\end{document}